\documentclass[trackchanges,twocolumn]{aastex701}
\usepackage[normalem]{ulem}
\usepackage{soul}

\newcommand{\soutPC}{\bgroup\markoverwith{\textcolor{cyan}{\rule[0.5ex]{2pt}{1pt}}}\ULon}

\usepackage{graphicx}
\usepackage{subcaption}

\usepackage{amsmath}
\usepackage{siunitx}
\usepackage[export]{adjustbox}
\newcommand{\comments}[1]{} 

\begin{document}
\title[Tidal disruption events with {\sc sph-exa}]{Tidal disruption events with {\sc sph-exa}: resolving the return of the stream}

\author[orcid=0009-0004-4569-6782,sname='Kubli']{Noah Kubli}
\affiliation{Department of Astrophysics, University of Zurich, Winterthurerstrasse 190, CH-8057 Z{\"u}rich, Switzerland}
\email[show]{noah.kubli@uzh.ch}  

\author[orcid=0000-0000-0000-0002,sname='Franchini']{Alessia Franchini}
\email[hide]{alessia.franchini@unimi.it}
\affiliation{Dipartimento di Fisica ``A. Pontremoli'', Università degli Studi di Milano, Via Giovanni Celoria 16, 20134 Milano, Italy}

\author[orcid=0000-0003-3765-6401,sname='Coughlin']{Eric R. Coughlin}
\email[hide]{ecoughli@syr.edu}
\affiliation{Department of Physics, Syracuse University, Syracuse, NY 13210, USA}

\author[orcid=0000-0002-2137-4146,sname='Nixon']{C. J. Nixon}
\email[hide]{C.J.Nixon@leeds.ac.uk}
\affiliation{School of Physics and Astronomy, Sir William Henry Bragg Building, Woodhouse Ln., University of Leeds, Leeds LS2 9JT, UK}

\author[orcid=0000-0003-3540-1405,sname='Keller']{Sebastian Keller}
\email[hide]{keller@cscs.ch}
\affiliation{Swiss National Supercomputing Centre, CSCS
Zurich, Andreasstrasse 5, CH-8092 Zürich, Switzerland}

\author[orcid=0000-0002-1786-963X,sname='Capelo']{Pedro R. Capelo}
\email[hide]{pcapelo@physik.uzh.ch}
\affiliation{Department of Astrophysics, University of Zurich, Winterthurerstrasse 190, CH-8057 Z{\"u}rich, Switzerland}

\author[orcid=0000-0002-7078-2074,sname='Mayer']{Lucio Mayer}
\email[hide]{lucio.mayer@uzh.ch}
\affiliation{Department of Astrophysics, University of Zurich, Winterthurerstrasse 190, CH-8057 Z{\"u}rich, Switzerland}



\begin{abstract}
In a tidal disruption event (TDE), a star is disrupted by the tidal field of a massive black hole, creating a debris stream that returns to the black hole, forms an accretion flow, and powers a luminous flare. Over the last few decades, several numerical studies have concluded that shock-induced dissipation occurs as the stream returns to pericenter (i.e., pre-self-intersection), resulting in efficient circularization of the debris. However, the efficacy of these shocks is the subject of intense debate. We present high-resolution simulations (up to $10^{10}$ particles) of the disruption of a solar-like star by a $10^6 \, M_{\odot}$ black hole with the new, GPU-based, smoothed-particle hydrodynamics (SPH) code {\sc sph-exa}, including the relativistic apsidal precession of the stellar debris orbits; our simulations run from initial disruption to the moment of stream self-intersection. With $\sim$$10^{8}$ particles -- corresponding to the highest-resolution SPH simulations of TDEs in the pre-existing literature -- we find significant, in-plane spreading of the debris as the stream returns through pericenter, in line with previous works that suggested this is a significant source of dissipation and luminous emission. However, with increasing resolution this effect is dramatically diminished, and with $10^{10}$ particles there is effectively no change between the incoming and the outgoing stream widths. Our results demonstrate that the paradigm of significant dissipation of kinetic energy during pericenter passage is incorrect, and instead it is likely that debris circularization is mediated by the originally proposed, stream-stream collision scenario.
\end{abstract}

\keywords{\uat{Tidal disruption}{1696}, 
\uat{Hydrodynamical simulations}{767},
\uat{Supermassive black holes}{1663}}


\section{Introduction}

Tidal disruption events (TDEs) involve the destruction of a star by the tidal field of a supermassive black hole (SMBH) and the subsequent accretion of tidally stripped debris \citep[][]{hills-75,lacy-82,rees-88,gezari-21}. The flares from these events are currently detected at a rate of $\sim$ tens per year \citep[e.g.,][]{vanvelzen21, hammerstein23,yao23, guolo24}, and this rate is expected to dramatically increase over the next few years due to facilities such as the Vera Rubin Observatory \citep[][]{ivezic19}.

TDEs sample a range of BH-related and physical processes, including circularization of material on elliptical orbits into a disk \citep[][]{rees-88}, accretion through both hydrodynamic and hydromagnetic instabilities \citep[e.g.,][]{sadowski-16}, accretion rate transitions from super- to sub-Eddington \citep[e.g.,][]{wu-18}, jet production \citep[e.g.,][]{Bloom2011}, and Lense-Thirring precession \citep[][]{stone12, franchini16, pasham24}. TDEs therefore present us with the possibility of establishing the properties of SMBHs in quiescent galaxies, and of developing our wider understanding of fundamental accretion physics.

That this remains a possibility (as opposed to a reality) is due in part to theoretical uncertainties in how the bound material transitions from highly eccentric orbits with eccentricity $e \sim 1$ to a (presumably still somewhat eccentric) disk that transports mass to the BH, i.e., we still do not have a complete physical picture of disk formation following a TDE. The original mechanism \citep[as put forward by][]{rees-88, evans-89} for facilitating the circularization of the gas in a TDE is the self-intersection of the incoming and outgoing debris streams (see Figure~\ref{fig-streamintersect} below). In this picture, the outgoing stream is largely thermodynamically unaltered during its pericenter passage, but is gravitationally deflected toward the incoming stream by relativistic apsidal precession; the small degree of differential apsidal precession across the stream near pericenter implies that it remains thin while doing so \citep[][]{evans-89, andalman-20}.

In addition to relativistic precession and the resulting stream collisions, \citet{evans-89,kochanek-94} discuss the possibility of shocks due to the compression of material upon returning to pericenter, highlighting that these may alter the debris velocities near pericenter and enhance circularization. Over the last few decades, independent numerical investigations -- using both finite-mass and finite-volume methods -- have found that the stream widens significantly upon passing through pericenter, with some of the (bound) material even being ejected on to unbound orbits \citep[e.g.,][]{lee96, ayal-00, guillochon-13, shiokawa15, ryu23, steinberg-24, price-24, Hu2025}. The origin of this effect has been largely attributed to the compression of the stream \citep[][]{evans-89,kochanek-94} and strong dissipation associated therewith; this is now commonly referred to as the ``nozzle shock'' \citep[see][for additional detailed discussion]{bonnerot-22}, which also provides an immediate source of emission and arguably obviates the self-intersection paradigm.

The plausibility of this newly emerging picture is, however, unclear because of the inherent difficulty in resolving the stream in numerical calculations as it returns to pericenter, which itself owes to the large spatial and temporal ranges encompassed by typical TDEs.\footnote{\label{footnote:1} The predicted apocenter distance of the most-bound debris from a solar-like star destroyed by a $10^6 \, M_{\odot}$ SMBH is $10^4 \, R_{\odot}$ with an orbital time of $\sim$30--40 days \citep[][]{rees-88}, compared to the sub-$R_{\odot}$ spatial scales and sub-stellar-dynamical temporal scales ($\lesssim 30$ minutes) necessary to resolve the original star; note that numerical simulations -- including those presented here -- find timescales that are shorter than this by a factor of the order unity (e.g., Figure 4 of \citealt{evans-89}, Figure 3 of \citealt{coughlin-15}).} Indeed, \citet{Huang2024} cast doubt on the veracity of simulations with strong ``nozzle shocks'', stating ``We note that we do not reach stream width convergence even given the highest resolution, and stream orbital plane expansion can still be related to numerical diffusion. The convergence study in \citet{price-24} also suggests that orbital plane expansion can be overestimated with insufficient resolution.'' This point can be seen by inspecting Figures 11 and 13 of \citet{Huang2024} and \citet{price-24}, respectively, which, as noted by \citet{Huang2024}, illustrate that the width of the reprocessed ``fan'' of debris, generated as the stream passes through pericenter, monotonically decreases with increasing resolution. \citet{bonnerot-22} similarly argued that the returning stream can be ``significantly affected by numerical artifacts caused by a too low resolution.''

That adequately modelling the return to pericenter of stellar debris is a difficult numerical problem is no surprise (see Footnote~\ref{footnote:1}). Even the original stellar disruption is difficult to resolve for sufficiently deep encounters (i.e., for high enough penetration factor $\beta = r_{\rm tidal}/r_{\rm p}$, where $r_{\rm p}$ is the orbit's pericenter distance and $r_{\rm tidal} = R_{\star}(M_{\bullet}/M_{\star})^{1/3}$ is the tidal radius, with $R_{\star}$, $M_{\bullet}$, and $M_{\star}$ being the stellar radius, BH mass, and stellar mass, respectively). For example, \citet{norman-21} showed that large spreads in the debris energies -- present in $\beta = 8$ and 16 encounters but absent at $\beta \lesssim 4$ -- largely disappear at sufficiently high resolution;\footnote{It is worth noting that, in this case, the results of the numerical simulations are substantiated by detailed analyses of the non-linear fluid dynamics as the star is vertically compressed near pericenter \citep[][]{Coughlin2022}.} if one identifies the effective $\beta$ of an encounter as the ratio of the distance from the SMBH at which the material stops being self-gravitating to the pericenter distance, then the stream from a solar-like star destroyed by a $10^6 \, M_{\odot}$ SMBH experiences $\beta \gtrsim 100$ \citep[a fluid element within the TDE stream stops being self-gravitating at effectively its Lagrangian apocenter;][]{coughlin-16}.

The numerical results discussed above and the reported trends with resolution lead to the obvious question: with sufficient accuracy, does the original picture envisaged by \citet{rees-88} -- a cold stream impacting a cold stream to drive circularization -- re-emerge? Here we show, with the highest-resolution simulations to date of the ``canonical TDE'' between a $5/3$-polytropic solar-like star destroyed by a $10^6 \, M_{\odot}$ SMBH, which are enabled by the highly efficient and highly scalable code {\sc sph-exa}, that the answer to this question is yes.

In Section~\ref{sec:methods}, we describe the numerical simulations and the initial conditions. In Section~\ref{sec:results}, we present the results of our study, and we conclude in Section~\ref{sec:conclusions}.

\section{Numerical Simulations}\label{sec:methods}

We use the novel code {\sc sph-exa} \citep[used, e.g., in][]{cabezon-25} to perform numerical simulations of the tidal disruption of a solar-like star by an SMBH. {\sc sph-exa} is a highly-efficient code that has been developed for use on graphics processing units (GPUs) with novel methodology and algorithms, which enables access to substantially higher-resolution simulations than has previously been possible. For details of the smoothed-particle hydrodynamics (SPH) implementation we refer the interested reader to \cite{cabezon-iad,cabezon-sphynx}, and for the specifics of the {\sc sph-exa} code to \cite{cabezon-25} and Appendix~\ref{code}. All of the simulations presented here were run on GH\,200 nodes of the ALPS supercomputer at the Swiss National Supercomputing Centre (CSCS). Despite the extreme geometric properties of this problem, which represent a huge numerical challenge, we achieved a performance of \SIrange{1e7}{2.5e7}{} particle timesteps per GPU per second.

Our simulations model the ``canonical TDE,'' which is composed of a solar-like star modelled as a polytrope with $M_{\star} = 1 \, M_{\odot}$ and $R_{\star} = 1 \, R_{\odot}$ with a polytropic exponent $\gamma=5/3$. TDE simulations are now typically performed with more realistic stellar density profiles \citep[][]{golightly-19-structure, goicovic-19, law-smith-20, jankovic-24}, and this is primarily done to understand the variation in the fallback rate as a function of these additional parameters. However, our main focus here is to understand the nature of the stream dynamics as the stream returns to pericenter, for which the standard polytropic setup is sufficient and which facilitates immediate comparison to earlier works (e.g., \citealt{evans-89, lodato-09, guillochon-13, Mainetti2017}).

To generate the initial star in {\sc sph-exa}, we begin with a template block composed of $50^3$ particles representing a glass-like constant-density system \citep[][]{arth-glass-19}. We then stack copies of this block in all dimensions until reaching the desired resolution $N$. This approach greatly reduces the initial noise compared to a random distribution of particles. The distribution is made spherical by discarding particles outside the largest fitting sphere. Through a radial coordinate transformation, we arrive at the desired density distribution corresponding to a polytrope in hydrostatic equilibrium. With this method, we can easily generate polytropes of arbitrary resolution, polytropic exponent, mass, and radius. We further relax the star (in isolation) by applying a damping force until an equilibrium is reached. 

We perform simulations varying the initial number of particles between $N = 10^6$ and $N = 10^{10}$, which spans and goes significantly beyond the highest-resolution SPH simulation of TDEs currently in the literature \citep[128\,M particles;][]{norman-21, fancher-23}. We place the star on a parabolic orbit around an SMBH at an initial distance of $5 \, r_\text{tidal}$ and pericenter distance $r_{\rm p} = r_{\rm tidal} = 100 R_{\odot}$. The SMBH is modelled with a pseudo-Newtonian potential that leads to the correct apsidal precession of the stream as it passes pericenter \citep[the Einstein potential;][]{nelson-papaloizou-2000}:

\begin{equation}
    \varPhi(r) = -\frac{GM_{\bullet}}{r}\left(1 + \frac{3r_\text{g}}{r}\right)\,,
    \label{eq-einstein}
\end{equation}

\noindent where $r$ is the spherical radius, $r_\text{g} = GM_{\bullet}/c^2$ is the gravitational radius of the SMBH, $G$ is the gravitational constant, and $c$ is the speed of light in vacuum. In all our runs, the SMBH has a mass of $M_{\bullet} = 10^6 \, M_{\odot}$. We also present in Appendix~\ref{code} the debris energy distributions from simulations that employ a Newtonian potential, for comparison with previous works.

We employ a polytropic equation of state with $P = K\rho^\gamma$, where $K$ is taken to be a global constant. This means that, while the gas temperature varies in response to adiabatic compression and expansion (i.e., $p{\rm d}V$ work), the heating of the gas due to dissipation (which occurs due to artificial viscosity\footnote{In under-resolved simulations, excess artificial viscosity can lead to excess numerical diffusion and dissipation. However, as the resolution is increased, the artificial viscosity applies the amount of dissipation necessary to correctly match the jump conditions for any shocks present in the flow. Thus, for converged simulations, the dissipation from the artificial viscosity is the dissipation introduced into the flow by any shocks that are present.}) is not included in the equation of state (and is lost from the system). A corollary of this approach is that the dissipation of kinetic energy by shocks is {\it overestimated}, because the compressing gas is not heated by artificial viscosity and is, thus, less able to withstand further compression. We do, however, keep track of the energy dissipated by the artificial viscosity terms, which serves as an upper bound on the physical dissipation due to shocks.

The star takes $\approx 3\,\text{hr}$ to reach pericenter, and we measure all times in the simulation with respect to this time of initial pericenter passage. The bound part of the resulting stream starts to return at $\approx 19\,\text{d}$, the analysis of which is the focus of the next section. 

\begin{figure*}
\centering
\begin{subfigure}{0.45\textwidth}
\centering
\includegraphics[width=1.0\textwidth]{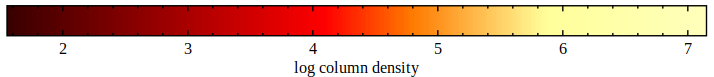}
\end{subfigure}

\begin{subfigure}{0.235\textwidth}

\begin{subfigure}{\textwidth}
    \centering
	\includegraphics[width=1.0\textwidth]%
    {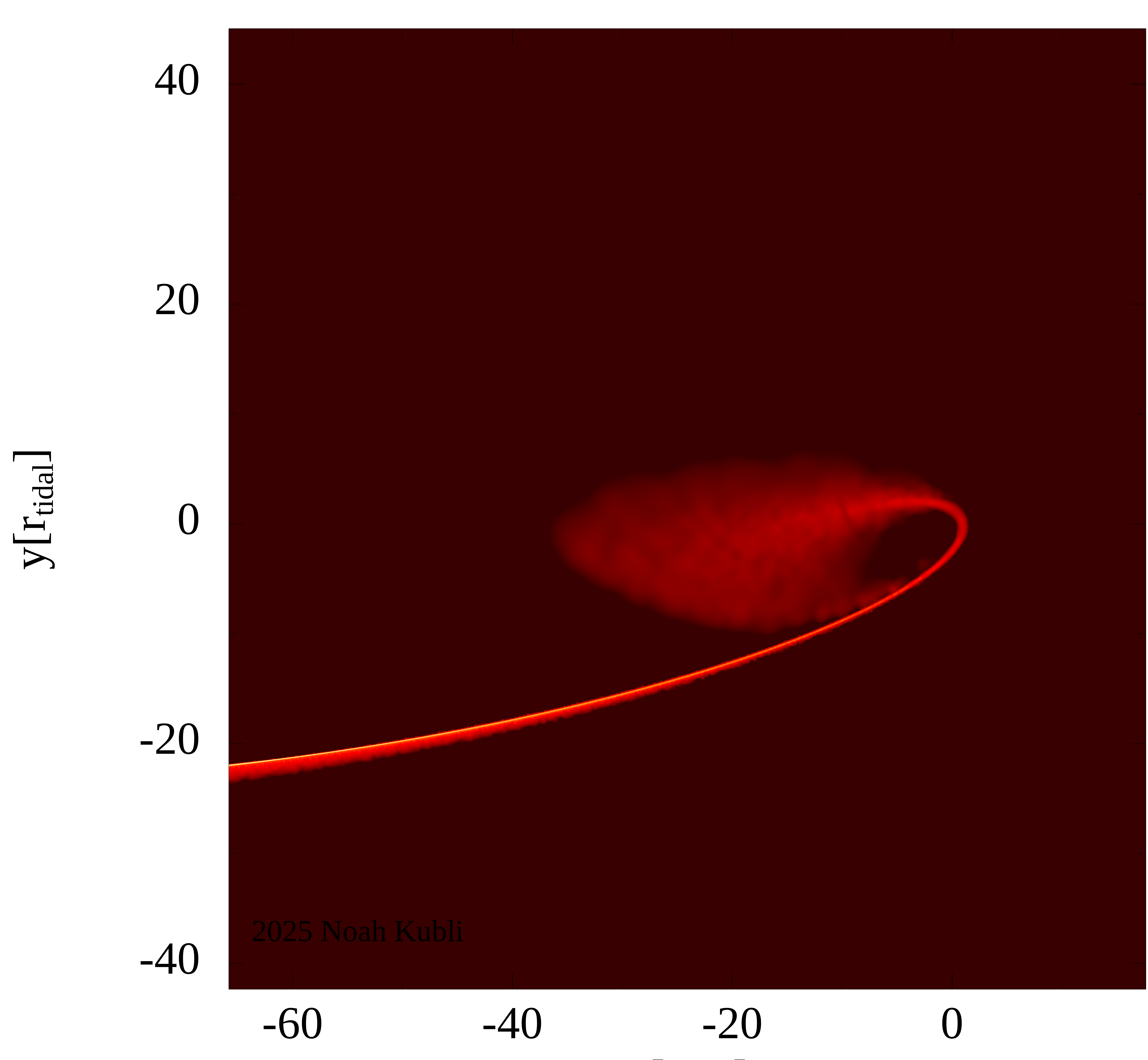}
\end{subfigure}
    
\begin{subfigure}{\textwidth}
	\centering
	\includegraphics[width=1.0\textwidth]%
    {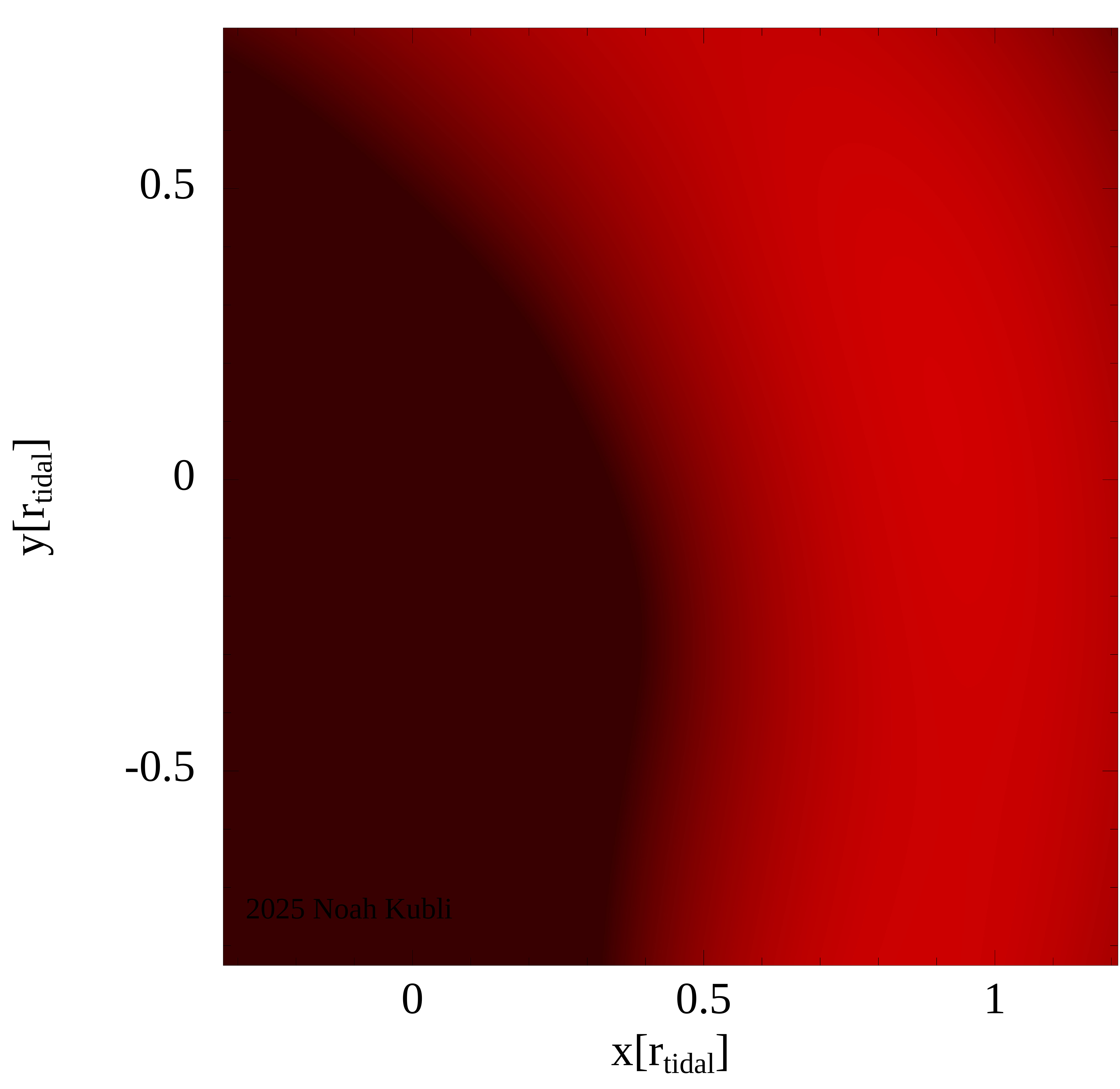}
\end{subfigure}

\begin{subfigure}{1.0\textwidth}
    \centering
    \includegraphics[width=1.0\textwidth]{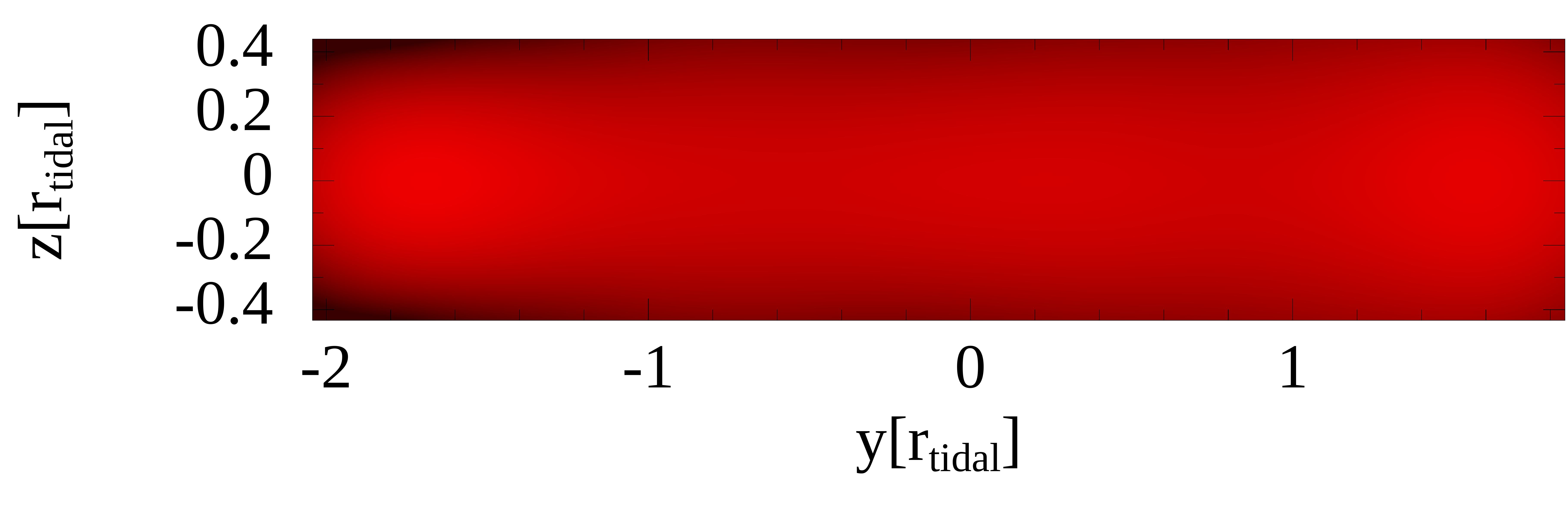}
\end{subfigure}

	\caption{1\,M}
\end{subfigure}%
\hfill
\begin{subfigure}{0.1912\textwidth}

\begin{subfigure}{\textwidth}
	\centering
	\includegraphics[width=1.0\textwidth]%
    {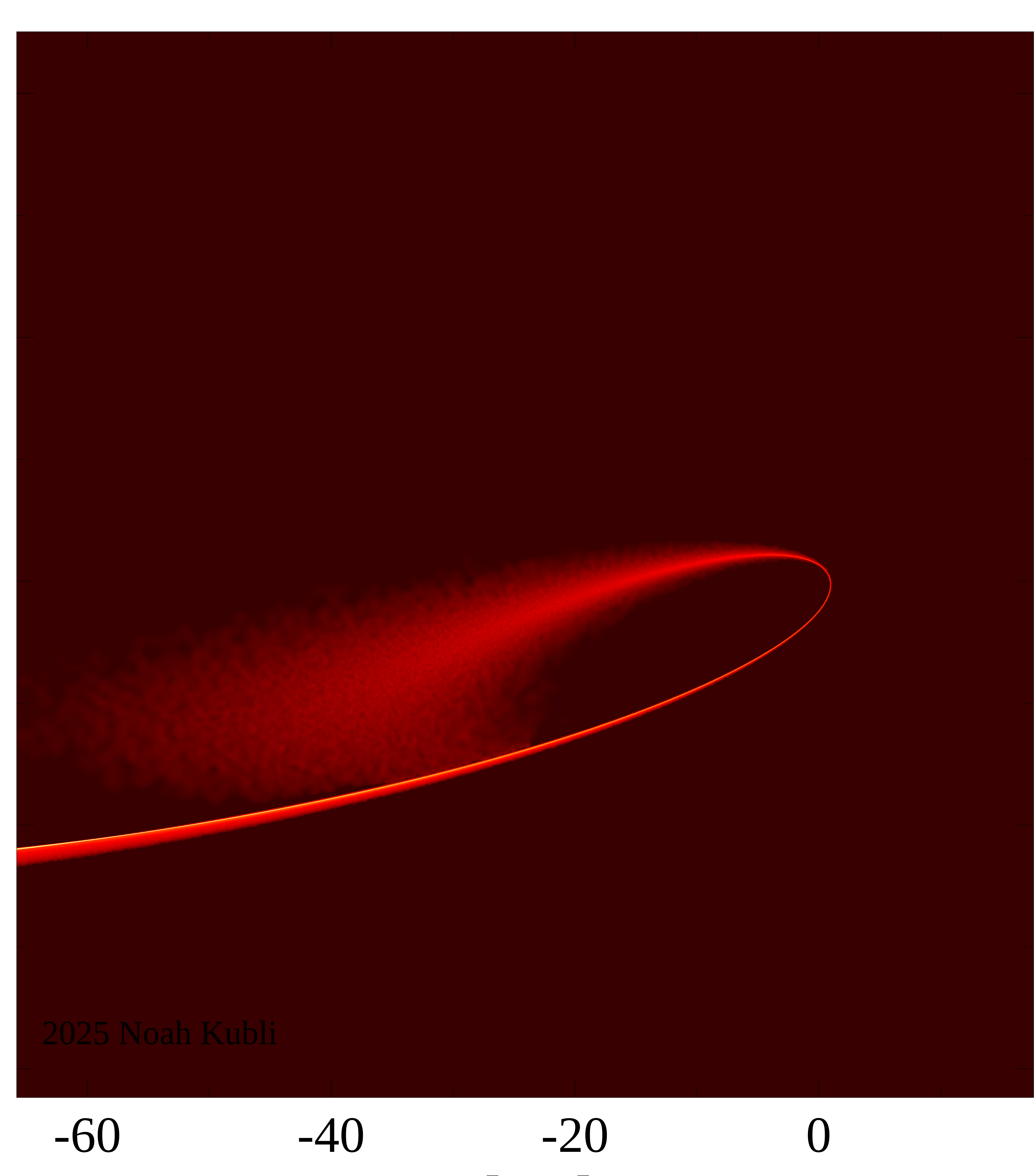}
\end{subfigure}

\begin{subfigure}{\textwidth}
	\centering
	\includegraphics[width=1.0\textwidth]%
    {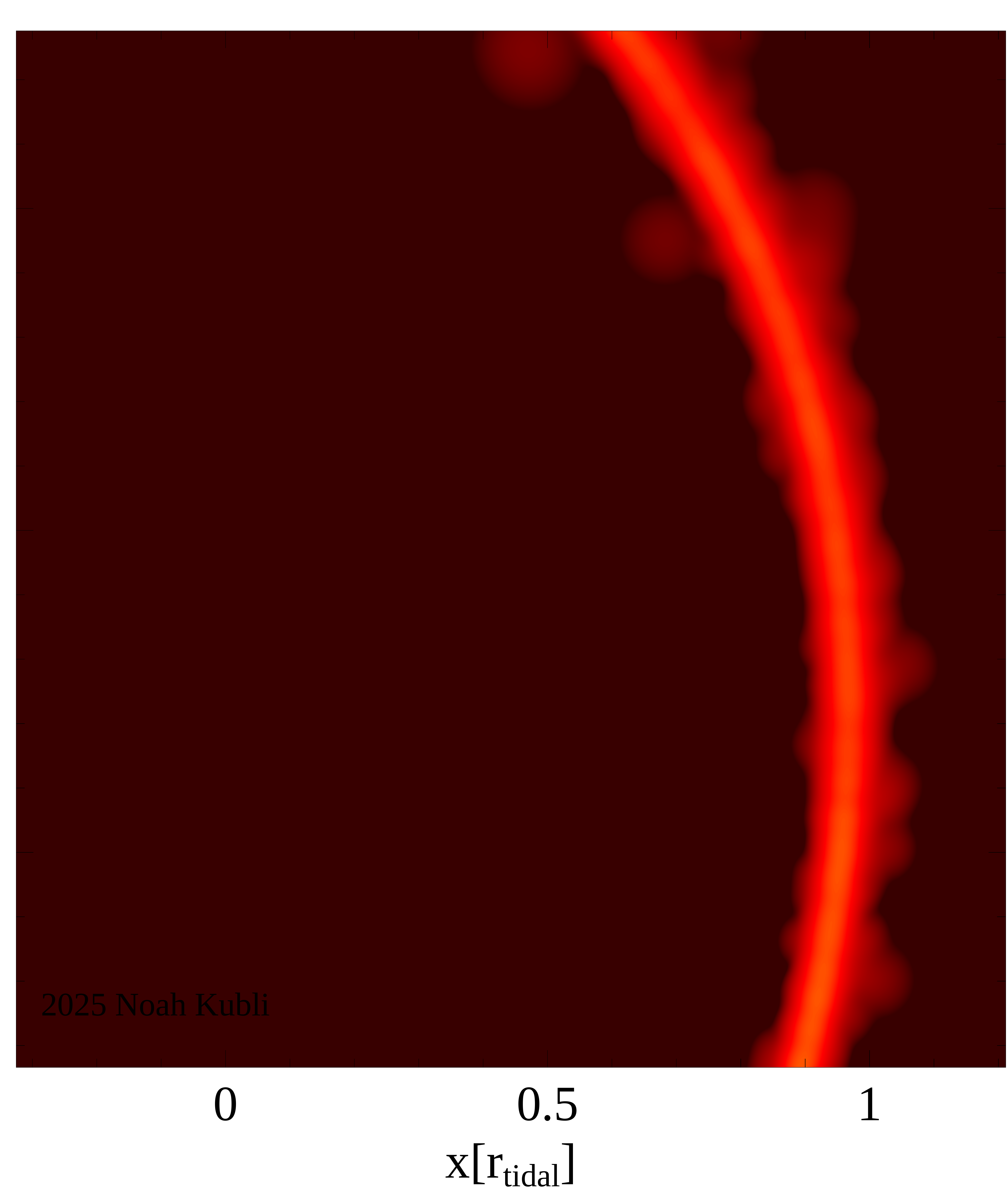}
\end{subfigure}

\begin{subfigure}{1.0\textwidth}
    \centering
    \includegraphics[width=1.0\textwidth]{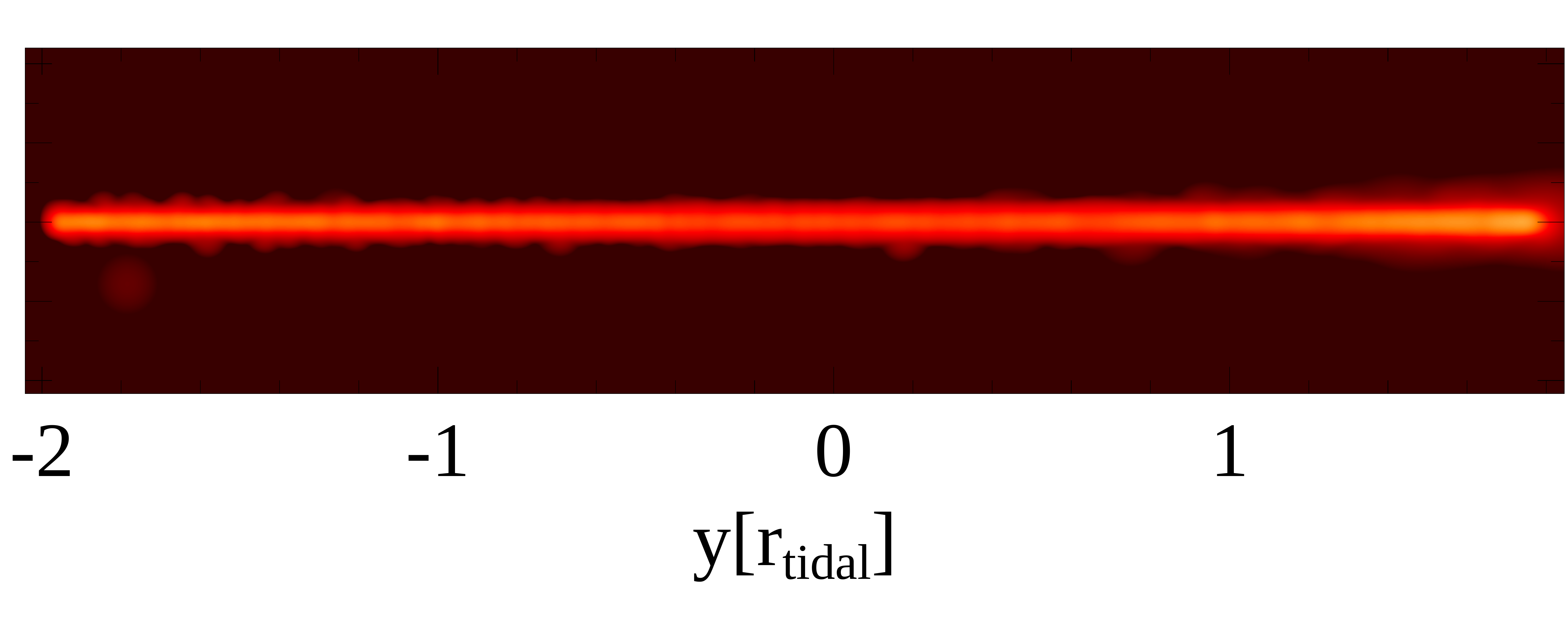}
\end{subfigure}
	\caption{16\,M}
\end{subfigure}%
\hfill
\begin{subfigure}{0.1912\textwidth}

\begin{subfigure}{\textwidth}
	\centering
	\includegraphics[width=1.0\textwidth]%
    {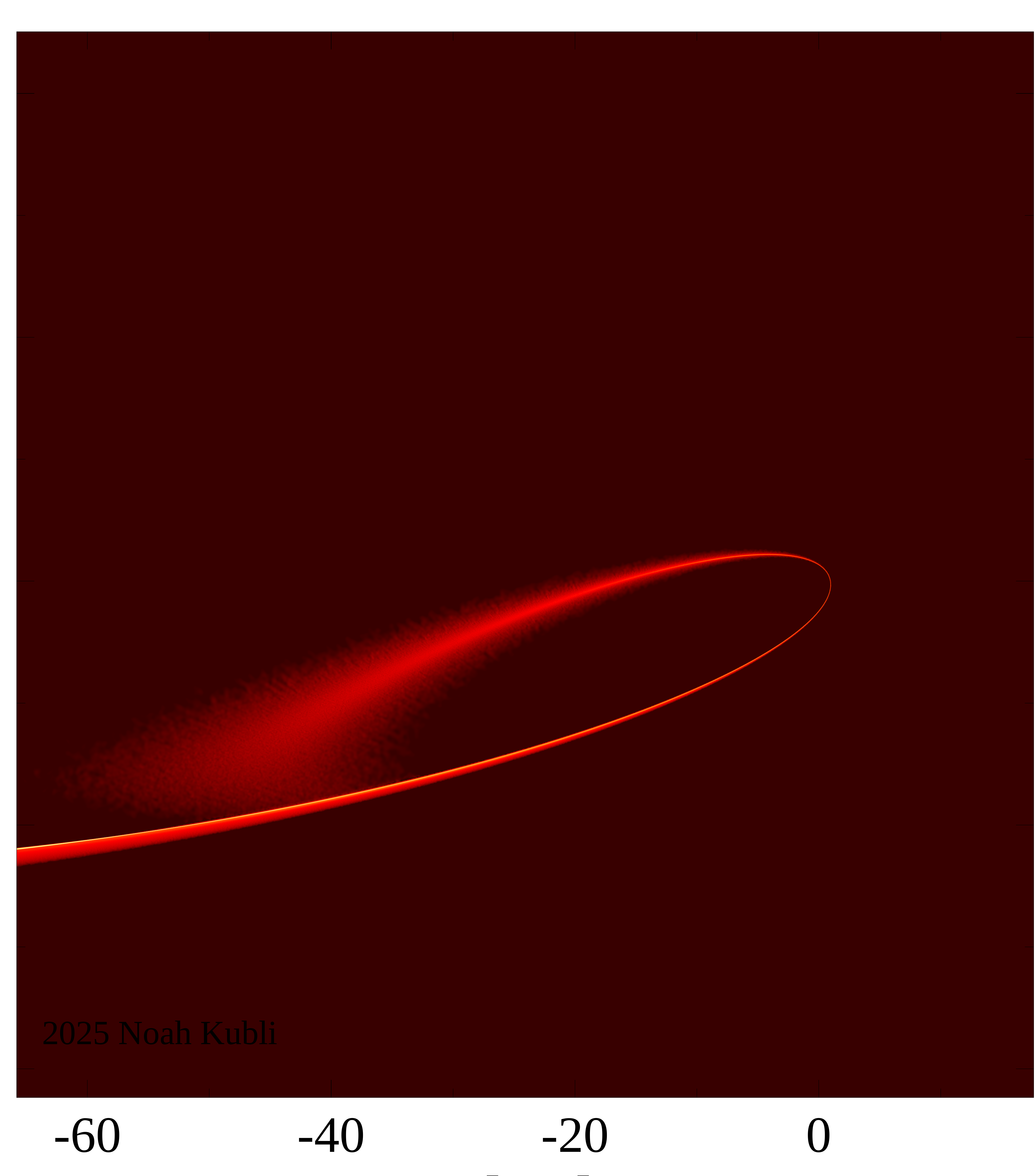}
\end{subfigure}

\begin{subfigure}{\textwidth}
	\centering
	\includegraphics[width=1.0\textwidth]%
    {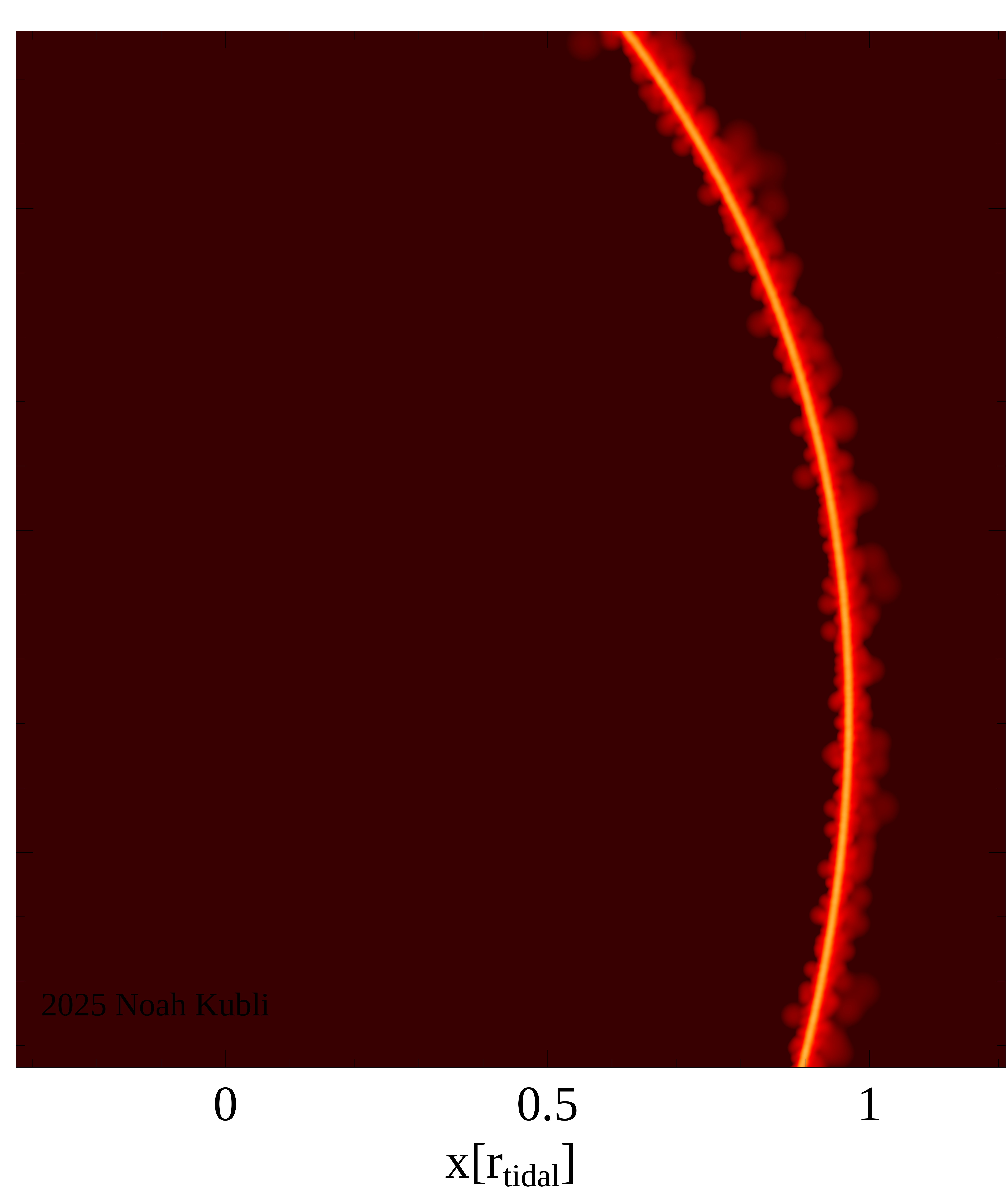}
\end{subfigure}

\begin{subfigure}{1.0\textwidth}
    \centering
    \includegraphics[width=1.0\textwidth]{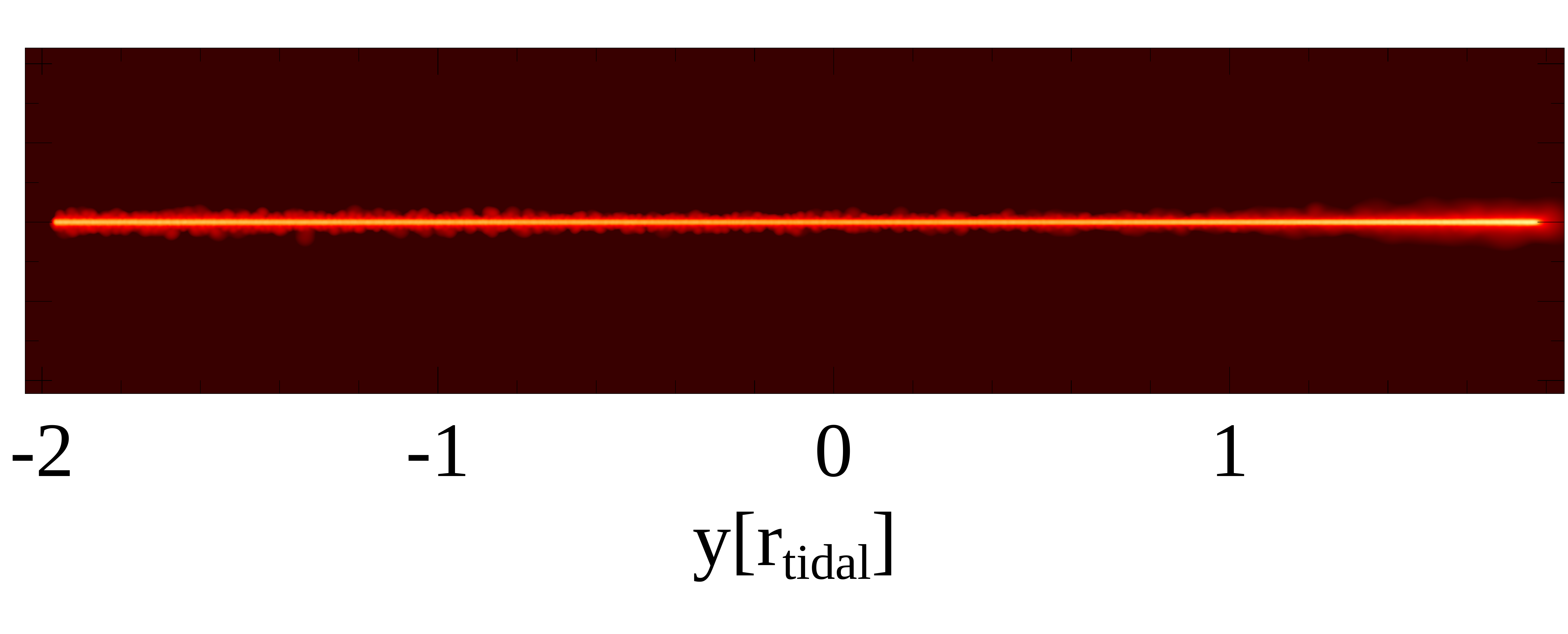}
\end{subfigure}
	\caption{128\,M}
\end{subfigure}%
\hfill
\begin{subfigure}{0.1912\textwidth}

\begin{subfigure}{\textwidth}
	\centering
	\includegraphics[width=1.0\textwidth]%
    {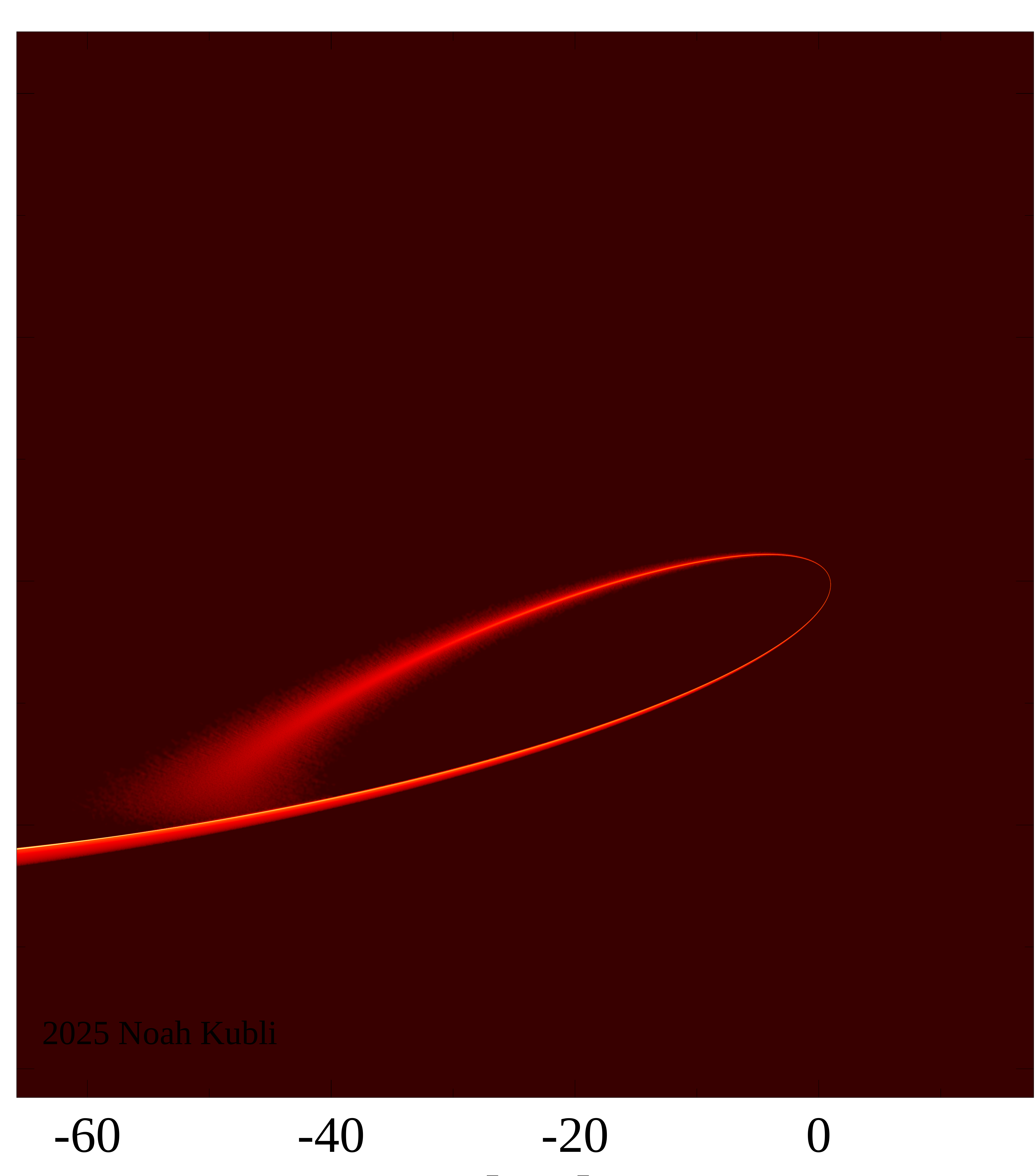}
\end{subfigure}

\begin{subfigure}{\textwidth}
	\centering
	\includegraphics[width=1.0\textwidth]%
    {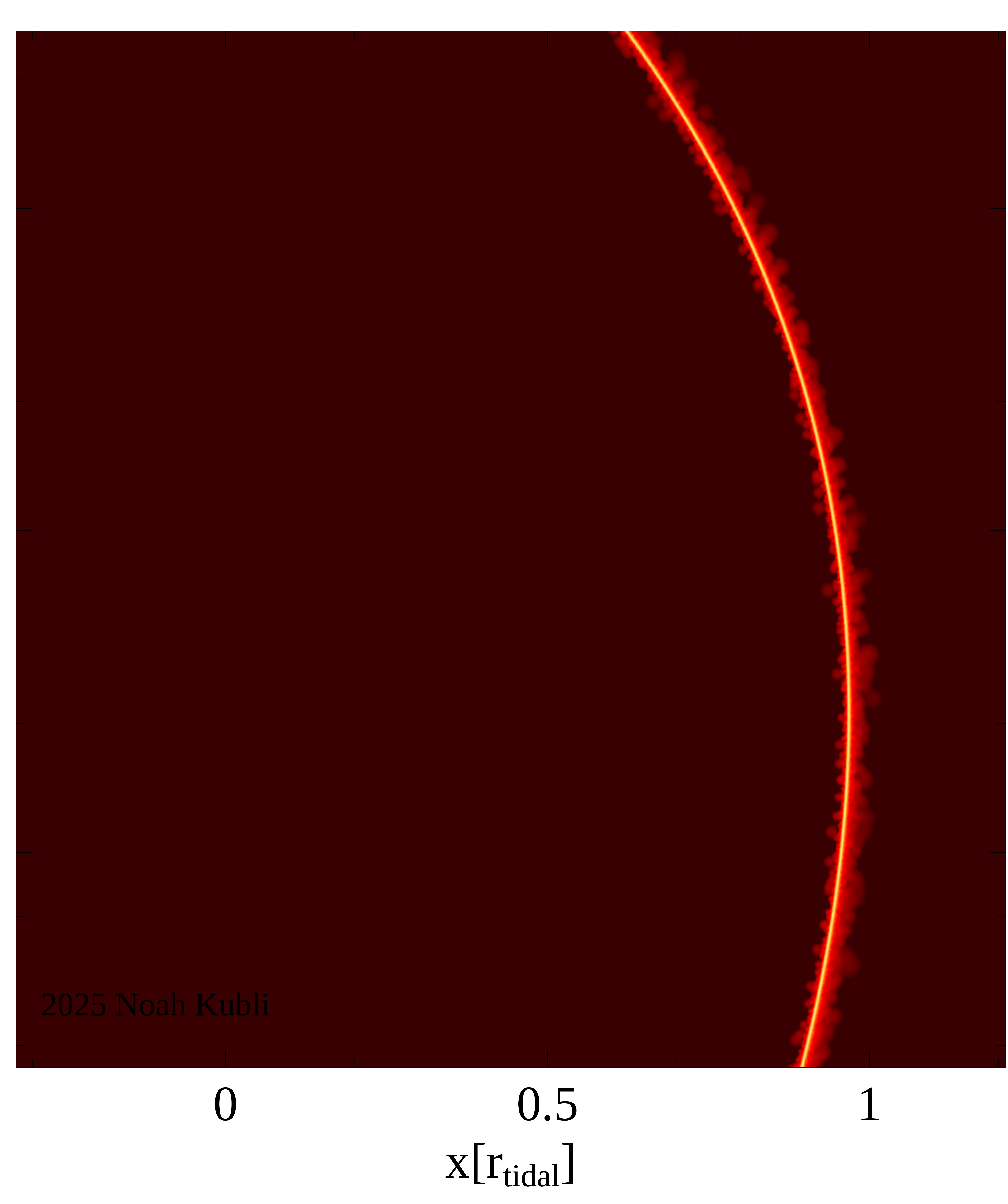}
\end{subfigure}

\begin{subfigure}{1.0\textwidth}
    \centering
    \includegraphics[width=1.0\textwidth]{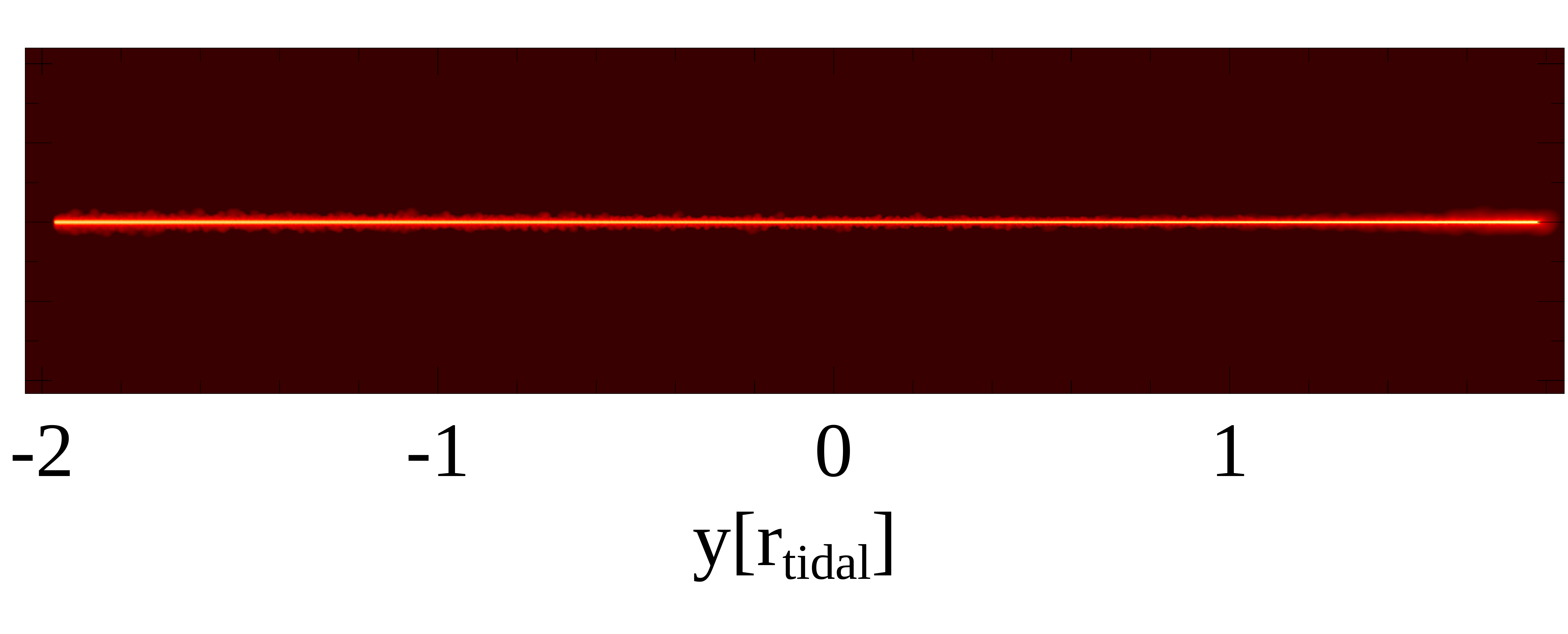}
\end{subfigure}
\caption{512\,M}
\end{subfigure}%
\hfill
\begin{subfigure}{0.1912\textwidth}

\begin{subfigure}{\textwidth}
	\centering
	\includegraphics[width=1.0\textwidth]%
    {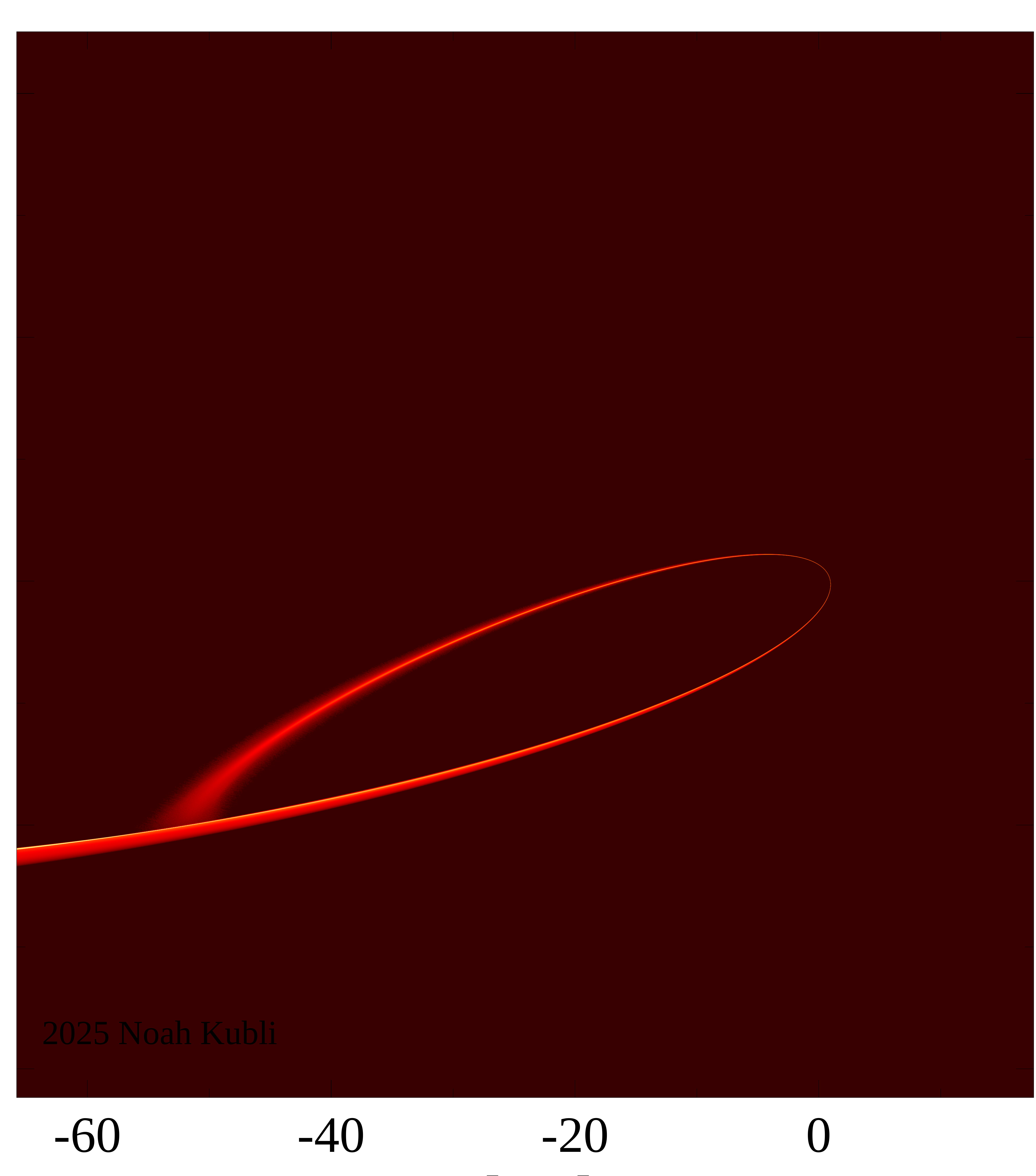}
\end{subfigure}

\begin{subfigure}{\textwidth}
	\centering
	\includegraphics[width=1.0\textwidth]%
    {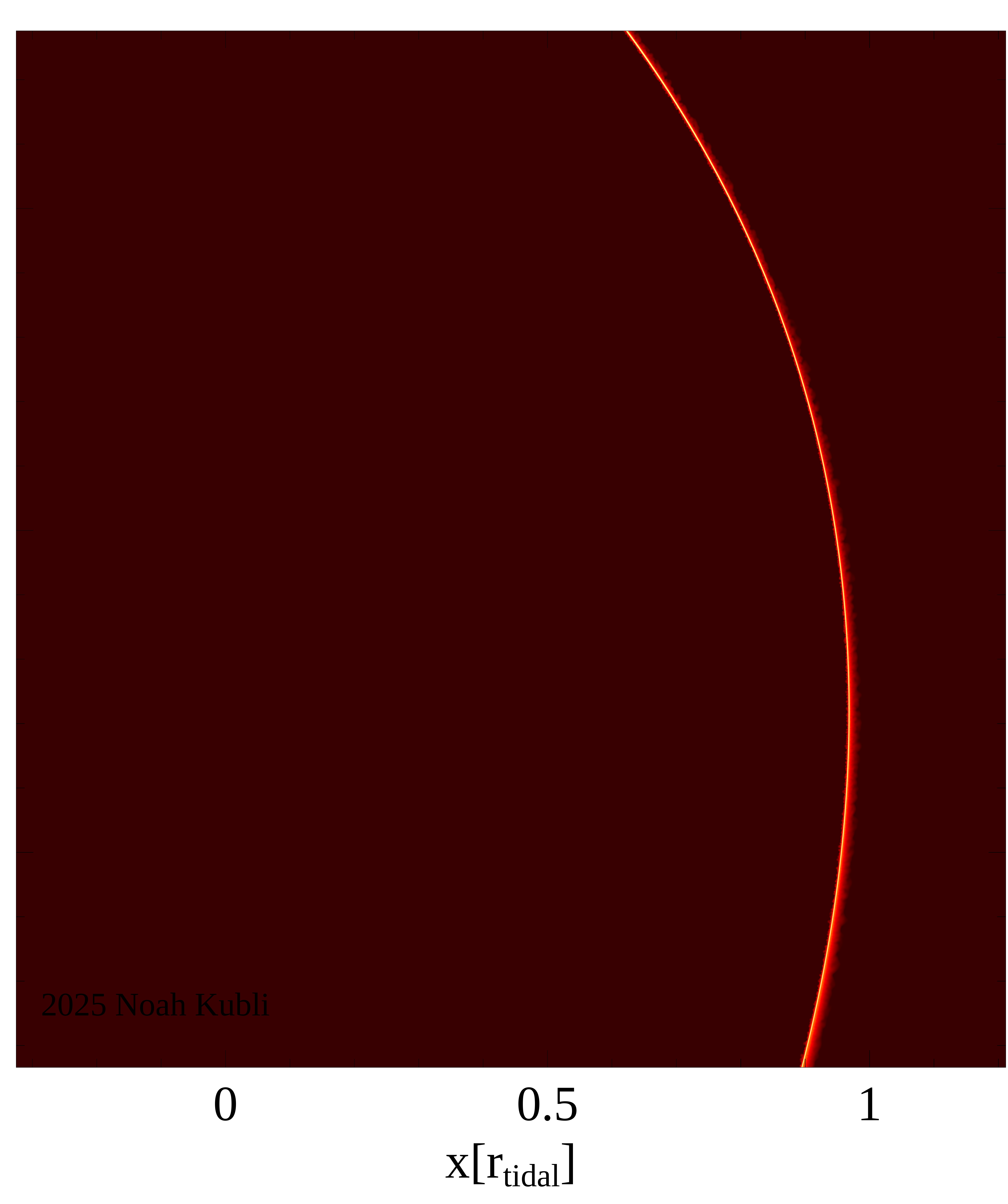}
\end{subfigure}

\begin{subfigure}{1.0\textwidth}
    \centering
    \includegraphics[width=1.0\textwidth]{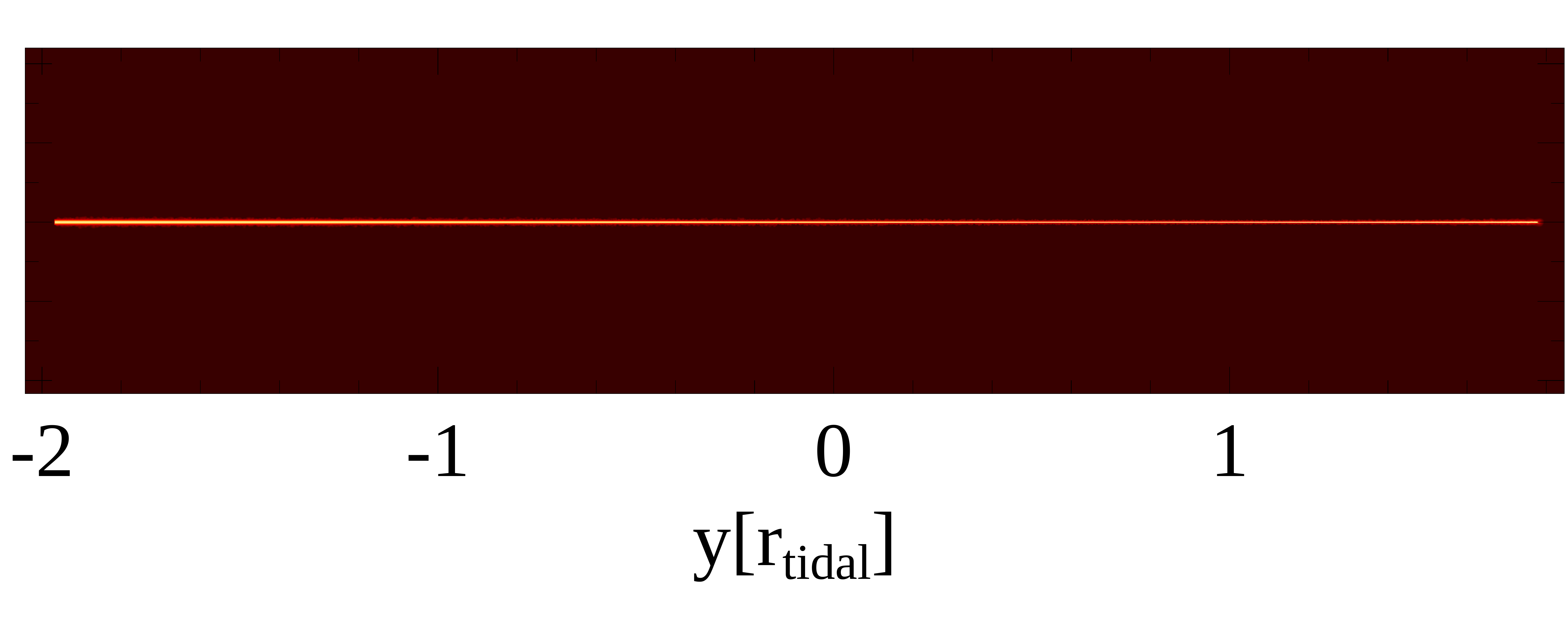}
\end{subfigure}
\caption{10\,B}
\end{subfigure}%
\hfill

\caption{Surface density maps of stellar debris stream at different resolutions at $t=26\,\text{d}$. The top panels show the region from pericenter to the location of the self-crossing of the stream due to apsidal precession, viewed face-on. The middle plots are zoomed-in versions of the top panels around pericenter. The bottom panels show an edge-on view on the stream close to pericenter. The surface density is in units of
$\text{g}~\text{cm}^{-2}$.}
\label{fig-streamintersect}
\end{figure*}

\section{Results}
\label{sec:results}

\begin{figure}
    \centering
    \includegraphics[width=0.5\textwidth]{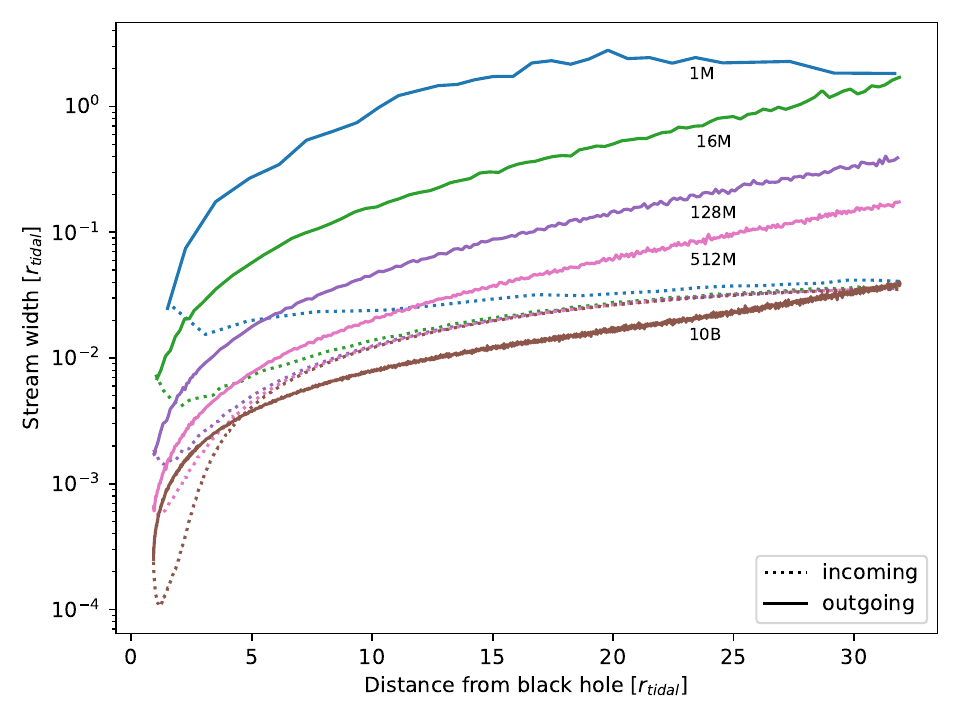}
    \caption{Widths of the incoming (dotted) and outgoing (solid) streams at $t=26\,\text{d}$, plotted over the distance from the BH in stellar tidal radii, using different resolutions.}
    \label{fig-stream-width}
\end{figure}

\begin{figure}
    \centering
	\includegraphics[width=0.5\textwidth]{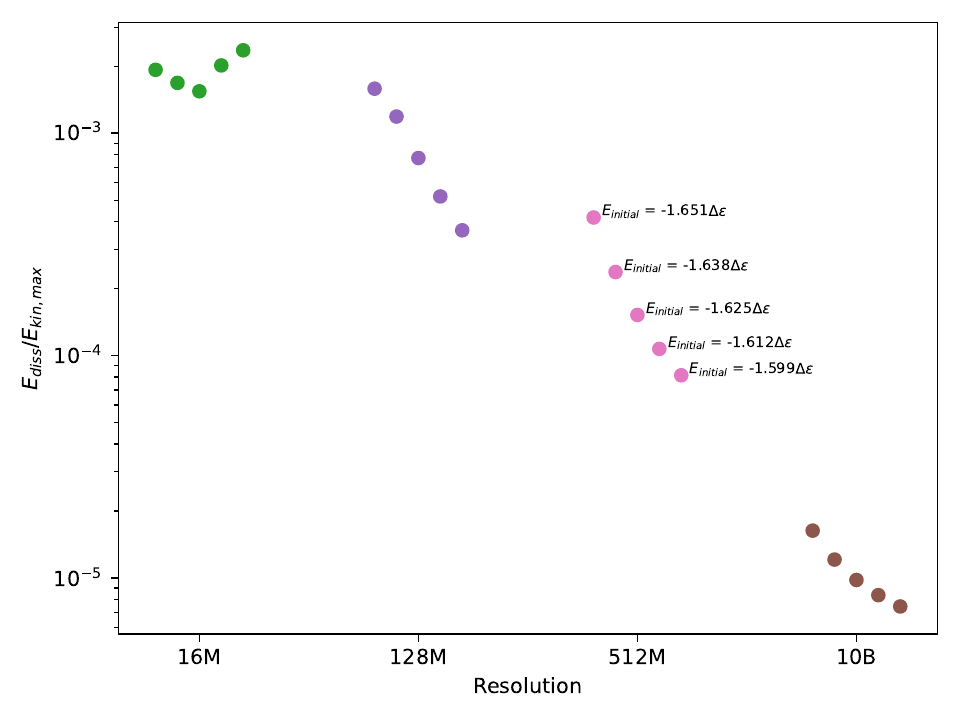}
    \caption{Energy dissipation at pericenter for different resolutions, measured at the time when the tip of the stream returns to pericenter at $t\approx19\,\text{d}$. Blue, orange, green, and red dots correspond to 16\,M, 128\,M, 512\,M, and 10\,B, respectively. We measure this for five parcels of matter at certain specific initial Keplerian orbital energies (annotated text, the same for all resolutions, $\Delta \epsilon = GM_{\bullet}R_{\star}/r_{\rm tidal}^2$). The dissipated energy is shown relative to the kinetic energy at pericenter.}
    \label{fig:energy_dissipation}
\end{figure}

Figure~\ref{fig-streamintersect} shows surface density maps of the region around the SMBH at $t=26\,\text{d}$, by which time the stream is just starting to self-intersect. There is an obvious and qualitative change in the outgoing stream width as a function of resolution: increasing the resolution from 1\,M to 10\,B particles leads to a significantly narrower stream, both near pericenter and closer to the self-intersection point. 

The width of the incoming and outgoing streams (at different resolutions) is shown in Figure \ref{fig-stream-width}. The width is measured by first classifying each particle as belonging to the incoming or the outgoing stream, depending on the sign of its radial velocity. We then divide both streams into radial bins, each bin containing the same number of particles, and in the $i^{\rm th}$ bin we compute the center of mass $r_{i}$. For each consecutive pair (outgoing and incoming) of $r_{i}$, we select all particles $\{j\}$ located between the planes perpendicular to the stream direction and passing through the respective points $r_{i}$ and $r_{i+1}$. We project the position of each particle in this subset onto the transverse direction $\{x^t_j\}$, which is orthogonal to the stream and in the $xy$-plane (i.e., perpendicular to the angular momentum vector of the original star's orbit), and compute the median absolute deviation as a statistically robust measure of the local stream width: $w = \text{med}_j(\lvert x^t_j - \text{med}_k(x^t_k)\rvert)$.

We also calculate the energy dissipated during the pericenter passage of the returning debris stream via the following algorithm: at $t\approx 18\,\text{d}$ -- when the most-bound region of the stream is at $\simeq 3 \, r_{\rm tidal}$ -- we select five adjacent parcels of gas within $r<30 \, r_\text{tidal}$ that are separated by a specific Keplerian orbital energy interval comparable to the initial binding energy of the star (at 10\,B particles, the most-bound parcel contains $10^5$ particles), and follow their thermodynamic evolution with time. Although we use a polytropic equation of state, we record the cumulative changes in the specific internal energy that arise from the pressure and artificial viscosity. We thus trace these particles through pericenter and determine the change in specific internal energy over this period. We then subtract the reversible (adiabatic) contribution to obtain the specific dissipated energy.

Figure~\ref{fig:energy_dissipation} shows the ratio of this dissipated energy to the kinetic energy at pericenter for four different resolutions, wherein the individual points correspond to the different gas parcels with different initial orbital energies (shown in the figure in units of $\Delta\epsilon = GM_{\bullet}R_{\star}/r_{\rm tidal}^2 = 100 GM_{\odot}/R_{\odot}$). The same set of initial orbital energies is used across all resolutions. The dissipated energy clearly decreases significantly with increasing resolution, dropping by more than two orders of magnitude between the 16\,M and 10\,B particle runs, and is $\sim$$10^{-5}$ of the kinetic energy at 10\,B particles. We reiterate that our polytropic equation of state overestimates the dissipated energy.

\section{Conclusions}\label{sec:conclusions}

Our simulations of the disruption of a solar-like star by an SMBH of mass $10^6 \, M_{\odot}$ -- performed at the highest resolution to date -- demonstrate that a) resolving the dynamics near pericenter requires {at least} $10^{10}$ particles, b) lower resolution results in unphysical ``spraying'' of the material near pericenter, and c) dissipation near pericenter (the ``nozzle'') amounts to at most $10^{-5}$ of the specific kinetic energy of the gas, and is thus irrelevant for circularization or the luminous output of TDEs. We therefore conclude that the post-pericenter spreading of the debris is not primarily driven by shock heating, but instead is largely a by-product of insufficient numerical resolution, in agreement with previous (albeit less direct) assessments by, e.g., \citet{bonnerot-22, Huang2024}.

Preceding works have claimed that dissipation near pericenter is responsible for the ejection of a considerable amount of mass \citep[][]{ayal-00} or the formation of ``winds'' and a large ``reprocessing envelope'' \citep[][]{price-24, hu25b}. A corollary of our findings is that such outcomes are not actualized by dissipation near pericenter alone, and (at least for the ``standard'' TDE that involves the disruption of a solar-like star by an SMBH) additional physics is required for producing these observationally relevant features, such as stream-stream collisions \citep[][]{jiang-16, Huang2024} leading to a quasi-spherical envelope of pressurized gas \citep[][]{metzger22} and/or super-Eddington feedback from the accretion flow \citep[][]{coughlin14, Meza2025}. 

Our conclusions have implications for the interpretation of observed TDEs. For example, if, as previously suggested, significant dissipation occurs for streams passing through pericenter, then one might expect all TDEs to behave in essentially the same manner; independent of the type of star, the nature of the disruption, or the properties of the BH, the debris is transformed into a quasi-spherical ball of hot gas. However, in the case that stream-stream collisions provide the dominant circularization mechanism, the resulting dynamics depends on the various properties of the system; e.g., the location of the collision depends on the pericenter distance of the initial stellar orbit and the mass (and spin) of the central BH. With the influx of data expected from, e.g., the Rubin Observatory, we might hope to see the effects of such parameters imprinted on the population(s) of observed TDEs. Additionally, it is clear that we must improve the physical models for debris stream dynamics, including, for example, the effects of recombination within the debris that can generate a thicker returning stream (\citealt{kochanek-94, coughlin23, steinberg-24}; Andalman et al., in prep.), potentially making simulations that resolve the stream more feasible.

\begin{acknowledgments}
We acknowledge useful correspondence with Chris Kochanek. ERC and CJN thank Zack Andalman and Eliot Quataert for useful discussions regarding the compression of the returning debris stream in TDEs. NK acknowledges support from the University of Zurich under the UZH Candoc Grant, \hbox{grant no. FK-25-093} and from the Swiss Platform for Advanced Scientific Computing (PASC) project SPH-EXA2. AF acknowledges financial support from the Unione europea - Next Generation EU, Missione 4 Componente 1 CUP G43C24002290001. ERC, CJN and LM thank the Kavli Institute for Theoretical Physics (KITP) where the project was initially conceived during the program ``Towards a Physical Understanding of Tidal Disruption Events,'' supported in part by grant NSF PHY-2309135 and the Gordon and Betty Moore Foundation Grant No. 2919.02 to KITP. ERC acknowledges support from the National Aeronautics and Space Administration NASA through the Astrophysics Theory Program grant 80NSSC24K0897. CJN acknowledges support from the Leverhulme Trust (grant No. RPG-2021-380). PRC acknowledges support from the Swiss National Science Foundation under the Sinergia Grant CRSII5\_213497 (GW-Learn). We also thank the Swiss National Supercomputing Center (CSCS) for their support as users of the ALPS supercomputer, on which the simulations were performed. In this work, we made use of {\sc splash} \citep[][]{price-splash-07} to create the rendered surface density plots.
\end{acknowledgments}

\section*{Data Availability}
The data underlying this article will be shared on reasonable request to the corresponding author.


\clearpage

\appendix

\section{The {\sc sph-exa} code}\label{code}

We use the novel code {\sc sph-exa} \citep[used, e.g., in][]{cabezon-25} for the simulations in this paper. {\sc sph-exa} couples SPH with a gravity solver, both running entirely on GPUs. The code is based on a new oct-tree implementation \citep[][]{keller_cornerstone_2023} optimized for distributed GPU machines, including an implementation of the Barnes-Hut algorithm \citep[][]{Barnes1986} for gravity and a neighbour search required for SPH. These algorithms, together with efficient node parallelism, allow for a massive speed-up compared to other codes, letting us run simulations at much higher resolution than before (or, for a given resolution, for a much longer timescale). The code runs on both NVIDIA and AMD GPUs and has shown efficient scaling on multiple supercomputers, including ALPS at CSCS in Switzerland and LUMI-G in Finland, using up to $2.8 \times 10^{12}$ particles.

The SPH implementation is based on \textsc{sphynx} \citep[][]{cabezon-sphynx}. When calculating the SPH gradients, we make use of the integral approach to derivatives \citep[IAD;][]{garcia-integral-sph,cabezon-iad} to increase the accuracy of SPH forces. In this method, the gradient of a function $f$ is computed as

\begin{equation}
    \nabla f_i = \sum_j V_j(f_j - f_i)A_{ij}\,,
\end{equation}

\noindent where $V_j = m_j / \rho_j$ is the volume element, with $m_j$ and $\rho_j$ being its mass and density, respectively, and

\begin{equation}
    A_{ij} = \boldsymbol{C}\cdot (\boldsymbol{x}_j - \boldsymbol{x}_i)W_{ij}(h_i)\,,
\end{equation}

\noindent with the coefficient matrix $\boldsymbol{C} = \boldsymbol{T}^{-1}$ and

\begin{equation}
\boldsymbol{T} = \sum_j V_j(\boldsymbol{x}_j - \boldsymbol{x}_i)(\boldsymbol{x}_j - \boldsymbol{x}_i)^T W_{ij}(h_i)\,,
\end{equation}

\noindent where $W$ is the kernel and $h$ is the smoothing length.

In our simulations, we use a sinc-kernel,

\begin{equation}
    W_{ij} = \frac{K}{h_i^3}
    \begin{cases}
        1 & q_{ij} = 0\,,\\
        \text{sinc}\left(\frac{\pi}{2} q_{ij}\right)^6 & 0 < q_{ij} \leq2\,,\\
        0 & q_{ij} > 2\,,
    \end{cases}
\end{equation}

\noindent where $K$ is the kernel normalization constant and $q_{ij} = \lvert \boldsymbol{x}_i - \boldsymbol{x}_j\rvert / h_i$, with 100 neighbours. The smoothing length $h_i$ is adaptive such that the neighbour count of each particle stays approximately constant over the simulation, and it is able to vary without an imposed ceiling or floor.

Artificial viscosity is added via

\begin{equation}
    \Pi_{ij}=
    \begin{cases}
        - v^\text{sig}_{ij}w_{ij}/2 & w_{ij}<0\,, \\
        0 & \text{else}\,,
    \end{cases}
\end{equation}

\noindent with $w_{ij} = (\boldsymbol{v}_i-\boldsymbol{v}_j)\cdot(\boldsymbol{x}_i-\boldsymbol{x}_j)/\lvert\boldsymbol{x}_i - \boldsymbol{x}_j\rvert$, $v^\text{sig}_{ij} = (c_i+c_j)/2 - 2 w_{ij}$, and $c_i$ being the speed of sound of particle $i$, computed from the (polytropic) equation of state. The high accuracy of the IAD approach, combined with the ability to reach extremely high resolutions by leveraging the largest parallel supercomputers, has enabled \textsc{sph-exa} to tackle successfully problems that are notoriously challenging for particle-based methods, such as subsonic turbulence, becoming competitive with state-of-the-art moving-mesh codes \citep[][]{cabezon-25}.

The gravity solver uses an opening angle criterion \citep[][]{Barnes1986} of $\theta_0 < 0.5$. When resorting to direct particle-to-particle calculations, gravity is softened and follows the smoothing length:

\begin{equation}
    \boldsymbol{F}_{ij} = G\frac{m_i m_j}{r_\text{eff}^3}
    (\boldsymbol{x}_j - \boldsymbol{x}_i)\,,
\end{equation}

\noindent where $r_\text{eff}^2 = \max(h_i^2, (\boldsymbol{x}_j - \boldsymbol{x}_i)^2)$.

We compute the integration timestep from two criteria. We apply the Courant condition \citep[][]{monaghan} to a particle $i$ as

\begin{equation}
    \Delta t_{\text{cour},i} =  K_\text{cour} \frac{h_i}{\max(0, \max_j\{c_i + c_j - 3 w_{ij}\})}\,,
\end{equation}

\noindent where the maximum is computed from the neighbors $j$ of particle $i$. An acceleration criterion is further applied as $\Delta t_{\text{acc}, i} = K_\text{acc}\sqrt{h_i/\lvert \mathbf{a}_i\rvert}$. We use $K_\text{cour} = K_\text{acc} = 0.2$ and employ the same global timestep for all particles, defined by the minimum of these two criteria.

\begin{figure}[h]
	\centering
    \begin{subfigure}{0.5\textwidth}
	\includegraphics[width=\textwidth
    ]{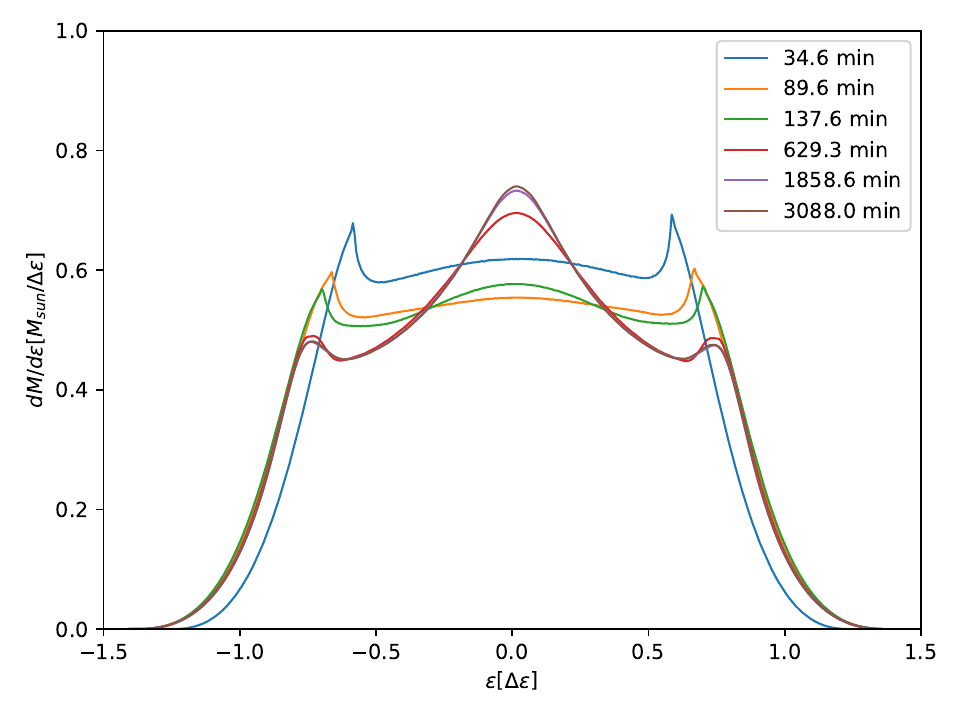}
    \end{subfigure}%
    \begin{subfigure}{0.5\textwidth}
    \includegraphics[width=\textwidth]{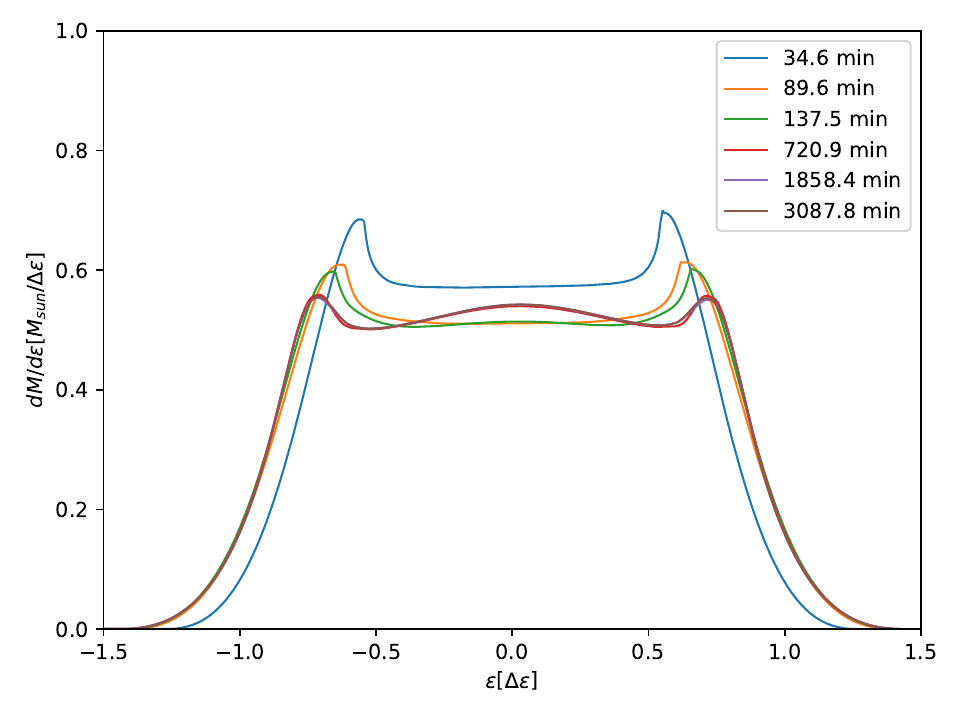}
    \end{subfigure}
\caption{Left-hand panel: specific energy distribution of the stellar stream after the initial disruption. This uses a resolution of 512\,M particles, and the SMBH is modelled as a Newtonian potential. Right-hand panel: The same measurement of a 10\,B particle simulation, modelling the SMBH with the Einstein potential. In both plots, we measure the time from the initial pericenter passage of the star onwards ($t=0$).}
\label{fig-energy-10B}
\end{figure}

As a benchmark for the accuracy of the code, we performed a set of TDE simulations that matched those performed by \citet{fancher-23}. Specifically, we ran simulations at 1\,M, 16\,M, 128\,M, and 512\,M particles (where 128\,M was their highest resolution) of the same disruption described in Section \ref{sec:methods}, but employing a point-mass Newtonian potential. Figure~\ref{fig-energy-10B} shows the distributions of debris energies at the times in the legend, which correspond (approximately) to the times in the legend of Figure 5 in \citet{fancher-23}. Comparing in the left-hand panel the curves at the respective times, we see that there are minor differences that arise on small energy scales (which have no bearing on the ability to resolve the returning stream near pericenter, i.e., the focus of the present work), which could be due to the different implementations of self-gravity and SPH forces between {\sc sph-exa} and {\sc phantom} \citep[][]{price-phantom}, which \citet{fancher-23} used for their simulations. However, the overall shapes and main features of the curves are in excellent agreement. We also show the debris energy distribution of our 10\,B particles simulation that includes the Einstein potential (Eq.~\ref{eq-einstein}) in the right-hand side of Figure~\ref{fig-energy-10B}.

\clearpage
\bibliography{bibliography}{}
\bibliographystyle{aasjournalv7}
\end{document}